\documentclass[twoside,11pt]{article}

\usepackage{jmlr2e}
\usepackage{hyperref}

\usepackage{amsmath}

\usepackage{algorithm}
\usepackage{algpseudocode}

\usepackage{color}
\definecolor{vermilion}{rgb}{1,0.3,0}
\definecolor{venetianred}{rgb}{0.78,0.03,0.08}
\definecolor{tyrianpurple}{rgb}{0.4,0.01,0.24}
\definecolor{forestgreen}{rgb}{0.13, 0.55, 0.13}

\usepackage{soul}

\newcommand{\ie}{\textit{i.e.}\ }

\newcommand{\aref}[1]{Appendix~\ref{#1}}
\newcommand{\eref}[1]{(\ref{#1})}
\newcommand{\fref}[1]{Figure~\ref{#1}}

\newcommand{\sref}[1]{Section~\ref{#1}}

\newcommand{\pkg}[1]{\textbf{#1}}

\newcommand{\rmd}{\mathrm{d}}
\newcommand{\rme}{\mathrm{e}}

\newcommand{\bx}{\boldsymbol{x}}

\newcommand{\bmu}{\boldsymbol{\mu}}
\def\bnu{\boldsymbol{\nu}}
\def\T{\mathrm{T}}
\def\mh{m^{h}}
\def\W{\mathcal{W}}
\def\am{\alpha_{\mu}}
\def\aw{\alpha_{w}}

\newcommand{\bbeta}{\boldsymbol{\beta}}
\newcommand{\btbeta}{\boldsymbol{\tilde{\beta}}}
\newcommand{\bX}{\boldsymbol{X}}
\newcommand{\bY}{\boldsymbol{Y}}
\newcommand{\bP}{\boldsymbol{P}}
\newcommand{\bp}{\boldsymbol{p}}

\newcommand{\N}{\mathcal{N}}

\ShortHeadings{The iBGe score for imperfect interventions}{Kuipers and Moffa}
\firstpageno{1}

\begin{document}

\title{The interventional Bayesian Gaussian equivalent score for Bayesian causal inference with unknown soft interventions}
\author{\name Jack Kuipers \email jack.kuipers@bsse.ethz.ch 
\\ \addr D-BSSE, ETH Zurich, Mattenstrasse 26, 4058 Basel, Switzerland
\AND 
\name Giusi Moffa \email giusi.moffa@unibas.ch \\ \addr Department of Mathematics and Computer Science, University of Basel, Basel, Switzerland \\ 
Division of Psychiatry, University College London, London, UK}

\editor{\hspace{-1.5cm}{\color{white}\rule[-0.2cm]{2cm}{0.6cm}} \vspace{-1.8cm}}

\maketitle

\begin{abstract}%   <- trailing '%' for backward compatibility of .sty file
Describing the causal relations governing a system is a fundamental task in many scientific fields, ideally addressed by experimental studies. However, obtaining data under intervention scenarios may not always be feasible, while discovering causal relations from purely observational data is notoriously challenging. In certain settings, such as genomics, we may have data from heterogeneous study conditions, with soft (partial) interventions only pertaining to a subset of the study variables, whose effects and targets are possibly unknown. Combining data from experimental and observational studies offers the opportunity to leverage both domains and improve on the identifiability of causal structures. To this end, we define the interventional BGe score for a mixture of observational and interventional data, where the targets and effects of intervention may be unknown. To demonstrate the approach we compare its performance to other state-of-the-art algorithms, both in simulations and data analysis applications. Prerogative of our method is that it takes a Bayesian perspective leading to a full characterisation of the posterior distribution of the DAG structures. Given a sample of DAGs one can also automatically derive full posterior distributions of the intervention effects. Consequently the method effectively captures the uncertainty both in the structure and the parameter estimates. Codes to reproduce the simulations and analyses are publicly available at \url{github.com/jackkuipers/iBGe}.
\end{abstract}


\section{Introduction}\label{intro}

Understanding and predicting the consequences of an action is the ultimate goal of investigation in many scientific disciplines. Questions about the effect of an intervention are of a causal nature and require understanding the causal relations between the variables under study. Directed acyclic graphs (DAGs) are convenient tools for representing causal mechanisms and help estimate intervention effects \citep{pearl95causal, greenland99causal, pearl00, bk:sgs00}. Gold standard methods for establishing the potential effect of an intervention rely on randomised studies. In reality, ethical, financial or practical difficulties may stand in the way of effectively and timely implementing experimental studies. In scenarios where trials are not an option but we have sufficient expert knowledge to draw a causal diagrams, we may use Pearl's do calculus \citep{pearl00} to evaluate the effect of potential interventions. In the absence of sufficient prior knowledge we need strategies that allow us to gain insights about a causal mechanism from observational data.

Causal discovery, however, relies on very strict assumptions, especially causal sufficiency. Furthermore, even under the assumption of no unmeasured confounders, observational data may only ever identify causal graphical structures up to a Markov equivalence class, also known as essential graphs \citep[EGs;][]{amp97} or completed partially DAGs \citep[CPDAGs;][]{chickering02CPDAG}. Methods limited to making inference on EGs may not be entirely satisfactory if we wish to fully characterise a causal mechanism. To resolve the uncertainty between equivalent DAGs we either need additional assumptions on the structural equations governing the relationships between variables, or we need to perform experiments to generate and collect interventional data. 

Since in practice it may only be possible to perform experiments on a subset of the variables in a domain of interest or a subset of the observational units, an appealing strategy is combining observational and interventional data to improve on the identifiability of causal structures. To extend existing Bayesian methods for structure learning and estimation of intervention effects to deal with a mix of observational and interventional data we need to define a (marginalised likelihood) score which accounts for the mixed nature of the data. Given a suitable score we can use recently developed methods \citep{ksm22} to efficiently sample from the posterior distribution of DAGs given the data. The procedure can provide a MAP (Maximum a Posteriori) estimator if of interest, but more importantly, it naturally accounts for the uncertainty in the structure learning task.

Early work to combine observational and interventional data for Bayesian structure learning appears in \cite{heckerman95} for deterministic interventions which was extended also to the more general case of non-deterministic manipulations in \cite{cooper1999causal}. Handling deterministic structural \citep[perfect or hard as per the definition in][]{eberhardt2007interventions} intervention is relatively straightforward, since the likelihood of the data of the intervened upon variables is simply 1 for the set value and we can ignore them when scoring each corresponding node as a child \citep{cooper1999causal}. When an intervened upon node acts as a parent, the intervention plays no role for the scoring of downstream nodes since the contribution to the score of each node is defined in terms of its conditional probability given the parents. The interventions, however, may disrupt the Gaussianity assumption used for example in the GIES \citep[Greedy Interventional Equivalence Search;][]{hauser2012characterization}; an algorithm developed to perform penalised maximum likelihood-based inference of causal structures from mixed observational and interventional data. An interesting line of developments is in the very recent work by \cite{castelletti2022network} presenting a Bayesian approach for Gaussian DAGs where the targets may be unknown but the interventions remain perfectly effective.

In many practical applications, such as genomic studies, imperfect interventions which only partially succeed (\ie soft interventions), or succeed a fraction of the time with a certain probability (\ie stochastic interventions), are not uncommon. To analyse data obtained under stochastic interventions, we can use a mixture model \citep{korb2004varieties} with a certain probability $\rho$ of a successful structural intervention severing all links into the intervened upon node and  the probability $(1-\rho)$ that the intervention is not successful and the unperturbed network still describes the data generating mechanism.

Deterministic soft interventions instead affect the relationship between a node and its parents in a common way across all observations obtained under the intervention. Additionally, both the strength and the targets of the intervention may be unknown. In a discrete data scenario, we can represent the interventions as additional nodes in the network (with no parents) and learn the targets by inferring their connections \citep{eaton2007exact}. In the soft interventional setting, representing the intervention as an additional parent amounts to modifying the relationship between the intervened upon node and the other parents (differently for each discrete parent state). The BDe score \citep{hg95} is fully parametrised (for a given set of parents including interventions) so it natively includes all interaction terms between the interventions and the other parents. This framework therefore provides a very general model of how an intervention may affect the node, covering the gamut from hard to soft interventions, though at the cost of increasing the parameter space. Learning the connections between the interventions and other nodes also allows for uncertainty in the targets \citep{eaton2007exact}. Although one may use alternative scores, the BDe scoring function has the advantage of automatically enabling a Bayesian approach.

Recent algorithms handling soft interventions, and also extended to deal with continuous data, include the general IGSP (Interventional Greedy Sparsest Permutation) algorithm of \cite{wang2017permutation} for known targets, a hybrid method with the score function defined in terms of conditional independence tests and structure learning with an order-based search, and the UT-IGSP (Unknown Target Interventional Greedy Sparsest Permutation) version with unknown targets \citep{squires2020permutation}.

With the current work we aim to bring the intrinsic flexibility of the discrete setting with the BDe score to scenarios with continuous data, by suitably adapting the BGe score \citep{gh02}. In particular, by leveraging the natural interpretation of interventions as interactions in the discrete setting and extending the strategy to continuous data we obtain a simple and powerful interventional BGe (iBGe) score, enabling scalable and accurate causal inference for continuous observational and interventional data in the presence of soft interventions with possibly unknown targets.

\section{The interventional BGe score}

\subsection{BGe score recap}

Consider an $n$-dimensional vector of random variables $\bX = \{X_1,\ldots,X_n\}$ and a dataset $d=\{\bx_1,\ldots \bx_N \}$ with $N$ observations of the vectors $\bx_i$. 
Under the model hypothesis $\mh$ that the distribution of $\bX$ is faithful to the DAG model, the marginal likelihood factorises into components for each node given its parents \citep{gh02}
\begin{eqnarray} \label{bgescore}
p(d \mid \mh) = \int p(d \mid \Theta, \mh) p(\Theta \mid \mh) \rmd \Theta &=& \prod_{i=1}^{n} \int p(d^{X_i} \mid d^{\bP_i}, \theta_i, \mh)  p(\theta_i \mid \mh) \rmd{\theta_i} \nonumber \\
& = &  \prod_{i=1}^{n} \frac{p(d^{X_i \cup \bP_i} \mid \mh)}{p(d^{\bP_i} \mid \mh)} 
\end{eqnarray}
where $d^{\bY}$ is the data restricted to the coordinates in $\bY\subseteq\bX$, $\Theta$ is the collection of all parameters in the model, $\bP_i$ are the parent variables of the vertex $i$ and $\theta_i$ are the parameters determining the conditional distribution of that node given its parents. While the first steps in \eref{bgescore} follow from the standard factorisation of a Bayesian network model and its likelihood into components of each node conditional on its parents, the final step to ensure the factorisation carries over to the marginal likelihood requires additional assumptions particularly on the prior distribution \citep{gh02}. For nominal categorical data, a multinomial dirichlet prior is required to meet all conditions and it leads to the BDe score \citep{hg95}. Joint Gaussian data require a normal-Wishart prior to satisfy all assumptions, leading to the BGe score which is the posterior probability of $\mh$, proportional to the marginal likelihood above and the prior on graphs.
 
For clarity we drop the explicit dependence on $\mh$ and since the score factorises we focus on a single node $X$ with parents $\bP$. The likelihood for the data $d^{X \cup \bP}$ consisting of $N$ observations $(x_i, \bp_i), i = 1\ldots N$ is
\begin{eqnarray} \label{pdatadef}
p(d^{X \cup \bP} \mid \bmu, W, \mh) &=& \frac{\vert W \vert^{\frac{N}{2}}}{(2\pi)^{\frac{(p+1)N}{2}}} \rme^{-\frac{1}{2}\sum_{i=1}^{N}\left[\bmu-(x_i, \bp_i)\right]^{\T} W \left[\bmu-(x_i, \bp_i)\right]}
\end{eqnarray}
following from the assumption of a jointly Gaussian distribution, with $W$ the precision matrix, $\bmu$ the mean and $p$ the number of parents.

By placing the conjugate Wishart prior on the full $n\times n$ precision matrix, $\tilde{W}\sim\W_{n}(T^{-1},\aw)$, where $\aw>n-1$ indicates the degrees of freedom and $T$ is the positive definite parametric matrix, and a normal prior on the full mean vector $\tilde{\bmu}$ with mean $\bnu$ and precision matrix $\am \tilde{W}$, with $\am>0$, the posterior distribution of $\tilde{W}$ and $\tilde{\bmu}$ are also normal-Wishart with updated parameters
\begin{align} \label{parameterupdate}
\am \to N +\am & &  \bnu \to \bnu' \nonumber \\
\aw \to  N +\aw & & T \to R 
\end{align}
where
\begin{equation}
\bnu' = \frac{N\bar{\bx} + \am \bnu}{(N+\am)} \, , \qquad R = T + S_N + \frac{N\am}{(N+\am)} \left(\bar{\bx}-\bnu\right)\left(\bar{\bx}-\bnu\right)^{\T}
\end{equation}
with 
\begin{equation}
\qquad \bar{\bx} = \frac{1}{N} \sum_{i=1}^{N} \bx_{i} \, , \qquad  S_N = \sum_{i=1}^{N}\left(\bx_i-\bar{\bx}\right)\left(\bx_i-\bar{\bx}\right)^{\T}
\end{equation}
as detailed in \cite{gh02, kmh14}. To compute the score for each node $X$, all that matters is the node itself and its parents with the contribution to the marginal likelihoods of
\begin{equation} \label{scoreratio}
\mathrm{BGe}(d, X) = \frac{p(d^{\bY} \mid \mh)}{p(d^{\bP} \mid \mh)} = \left(\frac{\am}{N+\am}\right)^{\frac{1}{2}}\frac{\Gamma\left(\frac{N+\aw-n+p+1}{2}\right)}{\pi^{\frac{N}{2}}\Gamma\left(\frac{\aw-n+p+1}{2}\right)} \frac{\vert T_{\bY\bY} \vert^{\frac{\aw-n+p+1}{2}}\vert R_{\bP\bP} \vert^{\frac{N+\aw-n+p}{2}}}{\vert T_{\bP\bP} \vert^{\frac{\aw-n+p}{2}}\vert R_{\bY\bY} \vert^{\frac{N+\aw-n+p+1}{2}}}
\end{equation}
where $\bY = X \cup \bP$, $\Gamma$ is the Gamma function and $A_{\bY\bY}$ means selecting the rows and columns corresponding to $\bY$ of a matrix $A$.

\subsection{SEM interpretation} \label{semversion}

Along with the matrix notation, we can reformulate the conditional distribution of $X$ on its parents $\bP$ in the Structural Equation Model (SEM) interpretation. If the matrix $B$ stores the edge weights of the DAG then the precision matrix is given by $\tilde{W} = (1-B) D (1-B)^{\T}$ where $D$ is a diagonal matrix of inverse variances. For the BGe score setting with a normal-Wishart prior on $\tilde{\bmu}$ and $\tilde{W}$, \cite{viinikka2020towards} explore in detail the expression of the posterior distribution of the edge weights deriving from the SEM reparametrisation and the consequent estimation of the causal effects.

In the absence of an intervention the structural equation at each node takes the form
\begin{equation}
X = \alpha + \bbeta \cdot \bP + \epsilon \, , \qquad \epsilon \sim \N(0, \sigma^{2})
\end{equation}
where $\bbeta = B_{\bP, X}$ and $\sigma^2 = D_{XX}^{-1}$. The likelihood for the observed data is simply the 1d Gaussian
\begin{equation}
p(d^{X} \mid d^{\bP}, \theta, \mh) = \frac{1}{(2\pi\sigma^2)^{\frac{N}{2}}} \rme^{-\frac{1}{2\sigma^2}\sum_{i=1}^{N}\left[x_i - \alpha - \bbeta\cdot\bp_i\right]^{2}}
\end{equation}
where $\theta$ collects the parameters $\alpha, \bbeta$ and $\sigma$. To compute the marginal likelihood and integrate over $\theta$ we can avoid the exact mapping to the normal-Wishart space by returning to the last step of \eref{bgescore}
\begin{eqnarray} \label{seminteqn}
\int p(d^{X} \mid d^{\bP}, \theta, \mh) p(\theta \mid \mh) \rmd(\theta) &=& \int \frac{1}{(2\pi\sigma^2)^{\frac{N}{2}}} \rme^{-\frac{1}{2\sigma^2}\sum_{i=1}^{N}\left[x_i - \alpha - \bbeta\cdot\bp_i\right]^{2}} p(\theta \mid \mh) \rmd(\theta)  \nonumber \\
&=& \frac{p(d^{X_i \cup \bP_i} \mid \mh)}{p(d^{\bP_i} \mid \mh)} = \mathrm{BGe}(d, X)
\end{eqnarray}
and utilising the simplification afforded by the prior choice which guarantees the factorisation of the marginal likelihoods.

\subsection{Hard interventions}

If in addition to a set of $N$ observations of $\bX$ we also have $N_{I}$ observations obtained after perfectly intervening on node $X$ and setting it to some normally sampled value stored in data $d_I$, the likelihood for the full data $\tilde{d} = (d, d_I)$ for node $X$ given its parents is
\begin{equation} 
p(\tilde{d}^{X} \mid \tilde{d}^{\bP}, \bmu, W, \mh) = p(d^{X} \mid d^{\bP}, \bmu, W, \mh) \prod_{i=N+1}^{N+N_I} f(x_i)
\end{equation}
where $f(x)$ is the interventional distribution \citep{hauser2012characterization}. When intervening, we break the connection between $X$ and its parents so that the data $d_I$ only contributes a constant factor to the likelihood and a constant term to the log-likelihood. Consequently there is no effect on the relative score of different DAGs. When removing the constant term corresponding to the interventional data, the formula above reduces to the same setting and result for the BGe score for the observational data $d$ alone. For scoring a node $X$ we simply remove all data where $X$ has undergone a deterministic hard intervention \citep{cooper1999causal} and then compute the BGe score as usual. Different nodes may undergo different interventions and the score for each node given its parents is only based on the data where that node has been observed under conditions without interventions.

\subsection{Soft interventions as interactions} \label{sec:onesoftint}

To consider soft interventions we return to the idea of \cite{eaton2007exact} of including them as additional parent nodes. In the discrete setting the scoring function automatically accounts for interactions between the intervention node and the other parents. For the continuous Gaussian case we wish to mimic the same structure, which naturally ensues with discrete data, and define the model in an analogous way. In particular given a mixture of observational and interventional data, we can view the intervention as another binary parent node $I$ and include an interaction term in the SEM
\begin{equation}
X = \alpha + \bbeta \cdot \bP + \tilde{\alpha}_{I} I + \btbeta_{I} \cdot \bP I + \epsilon(I)
\end{equation}
If we re-parameterise the regression coefficients ($\bbeta_{I} = \btbeta_{I} + \bbeta$, $\alpha_{I} = \tilde{\alpha}_{I} + \alpha$) we can rewrite the SEM as
\begin{equation}
X = \left\{ \begin{array}{lcl} \alpha + \bbeta \cdot \bP + \epsilon & & \text{for } I = 0 \\ \alpha_{I} + \bbeta_{I} \cdot \bP + \epsilon_{I} & & \text{for } I = 1 \end{array} \right.
\end{equation}
The above SEM representation implies that the conditional likelihoods of node $X$ in the observed data $d$ and intervened data $d_I$ takes the form
\begin{equation} \label{condprobint}
p(\tilde{d}^{X} \mid \tilde{d}^{\bP}, \theta, \mh) = \frac{1}{(2\pi\sigma^2)^{\frac{N}{2}}} \rme^{-\frac{1}{2\sigma^2}\sum_{i=1}^{N}\left[x_i - \alpha - \bbeta\cdot\bp_i\right]^{2}}\frac{1}{(2\pi\sigma_{I}^2)^{\frac{N_{I}}{2}}} \rme^{-\frac{1}{2\sigma_{I}^2}\sum_{i=N+1}^{N+N_{I}}\left[x_i - \alpha_{I} - \bbeta_{I}\cdot\bp_i\right]^{2}}
\end{equation}
In the case that the intervention may change all the parameters of node $X$ we can easily define the marginal likelihood contribution to the interventional BGe score
\begin{eqnarray} 
\mathrm{iBGe}(\tilde{d}, X) &=& \int p(\tilde{d}^{X} \mid \tilde{d}^{\bP}, \theta, \theta_{I} \mh) p(\theta, \theta_{I} \mid \mh) \rmd(\theta) \nonumber \\
&=& \int \frac{1}{(2\pi\sigma^2)^{\frac{N}{2}}} \rme^{-\frac{1}{2\sigma^2}\sum_{i=1}^{N}\left[x_i - \alpha - \bbeta\cdot\bp_i\right]^{2}} p(\theta \mid \mh) \rmd(\theta) \nonumber \\
& & \times \int \frac{1}{(2\pi\sigma_{I}^2)^{\frac{N_{I}}{2}}} \rme^{-\frac{1}{2\sigma_{I}^2}\sum_{i=N+1}^{N+N_{I}}\left[x_i - \alpha_{I} - \bbeta_{I}\cdot\bp_i\right]^{2}} p(\theta_{I} \mid \mh) \rmd(\theta_{I}) \nonumber \\
& = & \mathrm{BGe}(d, X)\times \mathrm{BGe}(d_{I}, X)
\end{eqnarray}
by applying \eref{seminteqn} to each term in \eref{condprobint}.

\subsection{Several interventions} \label{manyints}

A dataset may consist of observations from many different experimental conditions and several of them may affect the relationship between a node $X$ and its parents $\bP$. It is convenient to distinguish between interventions and experimental conditions which may consist of several interventions at the same time. For example an intervention could be a gene perturbation where some molecular agent targets the expression of a gene, while several such agents may be added together in a particular experimental setting. Since the effects of multiple soft interventions may not be simply additive we include potential interactions between the agents. Amongst a set of $m$ potential interventions $\{I_1, \ldots, I_{m}\}$ denote with $I_{X}$ the subset which are connected to the node $X$ along with the other observational parents $\bP$. Each combination of states of $I_X$ corresponds to a different experimental condition which we can equivalently represent by a categorical variable $E$ so that locally the SEM representation is
\begin{equation}
X = \left\{ \begin{array}{lcl} \alpha_0 + \bbeta_0 \cdot \bP + \epsilon_0 & &\text{for } E = 0 \\ 
\alpha_{1} + \bbeta_{1} \cdot \bP + \epsilon_{1} & & \text{for } E = 1 \\
\ldots & & \\
\alpha_{K} + \bbeta_{K} \cdot \bP + \epsilon_{K} & &\text{for } E = K
\end{array} \right.
\end{equation}
where there are $(K+1)$ distinct conditions. By including potential interactions between the effects of interventions this setting is a simple extension of the approach in \sref{sec:onesoftint}. By defining $\tilde{d}$ as the entirety of the data, and $d_k$ the subset of the data for which the experiment condition corresponds to category $k$, the marginal likelihood contribution to the interventional BGe score directly follows as
\begin{equation}\label{catiBGe}
\mathrm{iBGe}(\tilde{d}, X) = \prod_{k = 0}^{K} \mathrm{BGe}(d_{k}, X)
\end{equation}
For the full iBGe score, we multiply the marginal likelihoods above for each node $X$ and include the prior on graphical structures.

\subsection{Unknown targets}

The formula in \eref{catiBGe} allows us to score a DAG when the targets of each intervention are known. To extend to the case where the targets are unknown we also need to infer the edges between the intervention and the observation nodes \citep[akin to the discrete case,][]{eaton2007exact}. By implementing the interventional BGe score into the Bayesian sampling approach of \cite{km17, ksm22} we can learn and sample the structure as well as the targets of the interventions. One peculiarity is that the intervention nodes are fixed by the experimental setting and have essentially undergone hard interventions meaning that they have no parents in the network and can only affect downstream observables. The relevant size of the objective of inference is equal to the number $n$ of observed nodes, which corresponds to the size of the internal structure of the DAG excluding the intervention nodes.

\subsection{Causal effect estimation} \label{effectest}

Bayesian approaches can quantify not only the uncertainty in the network structure, but also the uncertainty in downstream analyses like the posterior distribution of interventional effects \citep{moffa2017using, kuipers2019links, moffa2021longitudinal}. The iBGe therefore also opens this possibility for mixed observational and interventional data in the linear Gaussian setting. For purely observational data, \cite{viinikka2020towards} derived the posterior distribution implied by the BGe score of the edge coefficients (the $\bbeta$ in the SEMs of \sref{semversion}) for each network. Combining DAG sampling with conditional parameter sampling given a structure we can build posterior distributions of causal effects and obtain a Monte Carlo estimate of hard (perfect) causal effects through the network.

For the iBGe case, since the interventions may be soft, only data generated in the natural state (the $d_0$ case of \sref{manyints}) for each node enter the computation of its edge coefficient estimates. Applying the formulae of \cite{viinikka2020towards} to the natural state data, we can sample the effects of perfect interventions for each network in the ensemble of structures drawn from the DAG posterior distribution to obtain the full posterior distribution of causal effects. In a linear setting, we can then obtain the distribution for known soft interventions with a simple weighted combination of the hard effects.   

\subsection{Software implementation}

To use our iBGe approach we interfaced the interventional BGe score with the \pkg{BiDAG} package \citep{skmb21} which implements a state-of-the-art hybrid method for structure learning and Bayesian sampling \citep{ksm22}
and which offers the highest performance for continuous observational data in benchmarking studies \citep{rmk21}. Along with defining the iBGe score, in the software implementation we treat the interventions as background nodes since they may have no parents in the network. For estimating causal effects, we also interfaced the implementation with the \pkg{Bestie} package (\url{https://CRAN.R-project.org/package=Bestie}). Code for computing the iBGe score, as well as for reproducing the simulations and the real data analysis is hosted at \url{github.com/jackkuipers/iBGe}.

\section{Simulation benchmarking}

As a proof of concept we first tested the performance of the BGe score in the well-understood case of hard interventions, with the results discussed in \aref{app:perfect_sims}. Here, we focus on the performance of the iBGe score in the more realistic case where both the targets of intervention, as well as the exact magnitude of their effects are unknown. As a relevant competitor handling unknown and soft interventions we include UT-IGSP (Unknown Target Interventional Greedy Sparsest Permutation) \citep{squires2020permutation}, an order-based greedy search with constraint-based tests on each order. The comparison excludes the recent Bayesian approach of \cite{castelletti2022network} since it assumes hard interventions even though the targets are unknown. Furthermore, it is a structure-based scheme which cannot scale to larger networks and it does not come with a software implementation.

\begin{figure}[t]
  \centering
  \includegraphics[width=\textwidth]{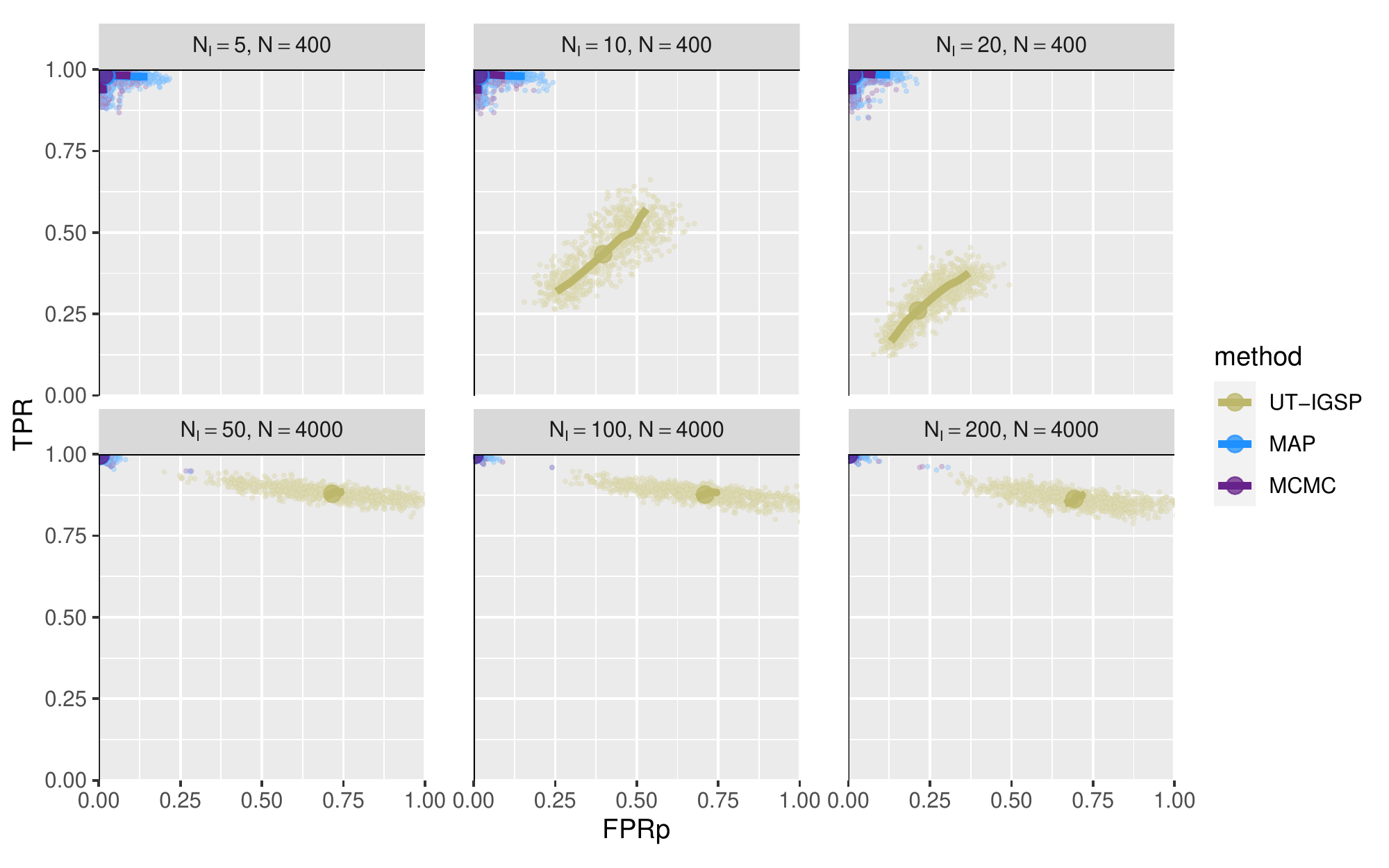}
  \caption{Comparison of the iBGe score input into the MAP and MCMC consensus scheme of \pkg{BiDAG} to UT-IGSP. Each point is a single repetition for a single parameter value, while the thicker lines show the average behaviour for each parameter value with the larger dot placed at  $10^{5}\alpha = 1 = 10\am$.}
  \label{fig:generalsim}
\end{figure}

\subsection{Simulation setting}

The simulation setup included data with $n=100$ observed variables and $m=10$ different interventions.  The graphical structure amongst the observed nodes was sampled as a random DAG with the default option of the \pkg{pcalg} package \citep{art:KalischMCMB2012} with an expected number of parents per node set to 2.  The number of targets of each intervention was sampled from a Poisson with rate parameter 1 shifted by 1, while the targets themselves were sampled uniformly from the 100 observed nodes.  The edge weights in the DAG were sampled uniformly in the range $[0.25, 1]$. The interventions were non-overlapping and their effect on the target nodes was to shift their mean by an amount sampled from a standard normal, and to damp the effects of the other parents by multiplying their edge weights by a uniformly sampled number from the interval $[0.1, 1]$. The data was generated from the SEM in topological order, where each node is given by the linear combination of its parents in the DAG (possibly modified by the interventions) and with standard normal noise added on top. Although the iBGe score and other score-equivalent approaches are insensitive to the scale of the data, other metrics may not be \citep{reisach2021beware}. To avoid any scale-related artefacts we standardise the data by default.

For each of the 10 interventions we generated $N_I = (5, 10, 20)$ observations under that condition, and then added purely observational data to achieve $N=400$ observations in total. In addition we examine a large data setting where the number of each type of observation is multiplied by 10. To capture the sampling variability we repeated the simulation of each setting 100 times.

\subsection{Performance measure} \label{sec:perf}

As a measure of performance we compared the inferred DAG to the data-generating DAG, after mapping both to the equivalence-class space. Although the inference schemes are unaware of the actual targets, they are used for deriving equivalence classes for the performance evaluation \citep{hauser2012characterization}. We compute the number of TP edges (directed edges in the same direction in inferred and model graphs, or undirected edges in both) and the number of FP edges (directed or undirected edges in the inferred graph not in the model graph). Edge directions which do not match (wrong direction or undirected) between the inferred and model graphs count as $\frac{1}{2}$ to FP and $\frac{1}{2}$ to FN so that the structural Hamming distance (SHD) is
\begin{equation}
\mathrm{SHD} = \mathrm{FN} + \mathrm{FP} = \mathrm{P} - \mathrm{TP} + \mathrm{FP}
\end{equation}
and it coincides with the Manhattan distance from $(0, \mathrm{P})$ in a TP vs FP plot. Since the total number of edges is random, we scale by P and plot the TPR against FPRp = $\frac{\mathrm{FP}}{\mathrm{P}}$. The performance further depends on algorithmic parameters, so we create ROC-like curves by varying the significance level $\alpha$ of the independence tests in UT-IGSP and by varying the prior parameter $\am$ in the iBGe score while keeping $\aw = \am + n + 1$. For the plots we used the following values
\begin{equation}
10^{5}\alpha = (0.00248, 0.0111, 0.0498, 0.223, 1, 1.65, 2.72, 4.48, 7.39) = 10\am
\end{equation}
\begin{figure}[t]
  \centering
  \includegraphics[width=\textwidth]{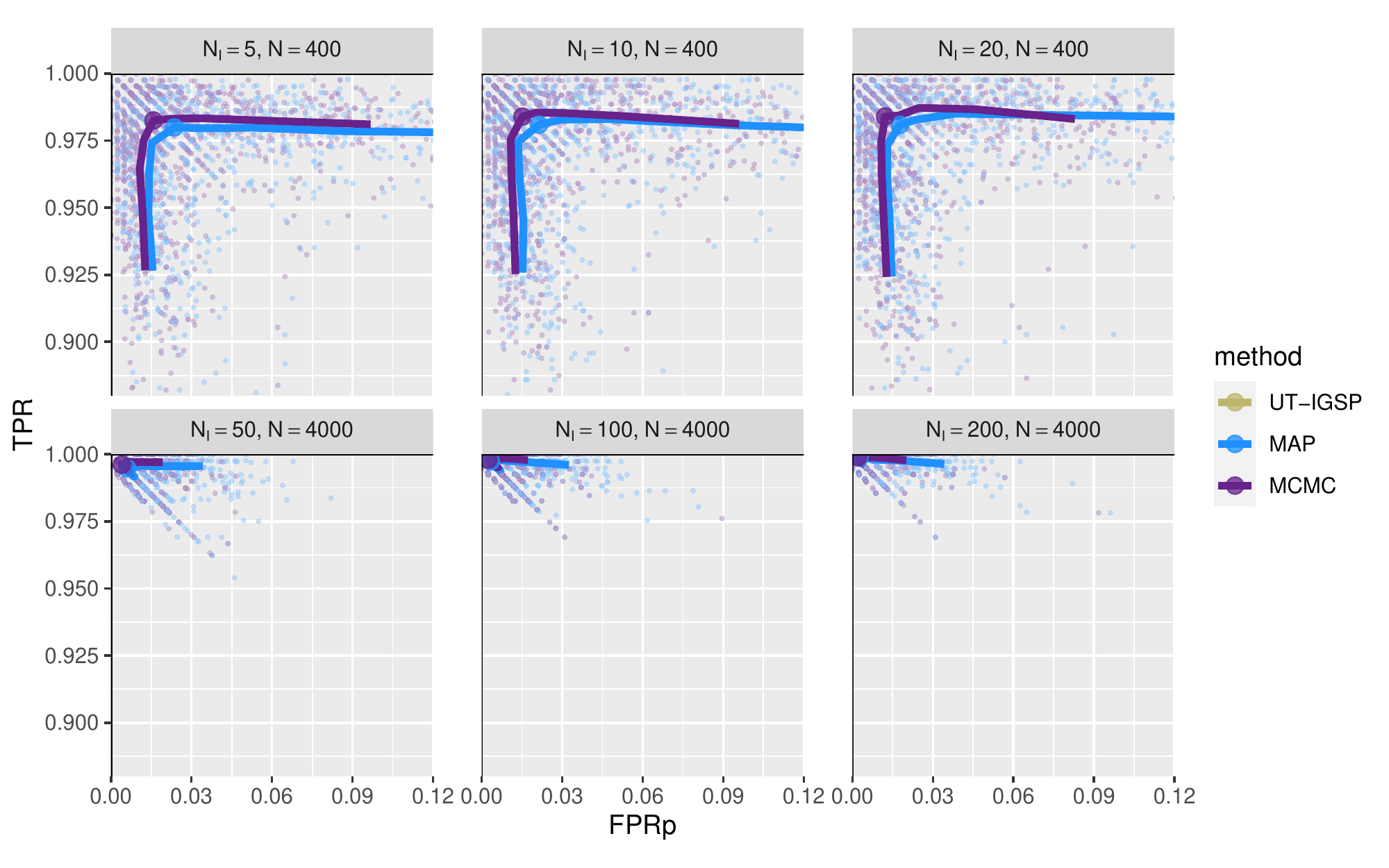}
  \caption{We zoom into the top left of \fref{fig:generalsim} to examine how closely our MAP and MCMC methods with the iBGe score approach (0,1). At the smaller sample sizes, the differences are typically a few edges with a slight advantage from the Bayesian model averaging with the MCMC scheme. At the larger sample size, we often have perfect performance.}
  \label{fig:generalsim_zoom}
\end{figure}

\subsection{Simulation results}

Employing the iBGe score for the iterative MAP search with the \pkg{BiDAG} package, and building a consensus graph through posterior thresholding (with threshold 0.5) from a sample of DAGs using its MCMC scheme achieves very high performance (\fref{fig:generalsim}). The constraint-based UT-IGSP algorithm appears to perform quite poorly. For the lowest number of samples per intervention of $N_I=5$ with a total of $N=400$ samples, the algorithm runs into numerical errors and cannot complete when it tries to build and test the covariance matrix in each experimental condition. With twice as many samples per intervention ($N_I=10$), UT-IGSP typically fails to find half the true edges in the graph, and this gets worse with more interventional samples as this setting has fewer observational samples (with the fixed total of 400). Increasing the sample sizes by a factor of 10 (\fref{fig:generalsim}, bottom row), drastically improves the number of true edges found by UT-IGSP, but also leads to large numbers of false positives. When comparing the bottom row of \fref{fig:generalsim} with the top row it is apparent that the UT-IGSP even with much larger sample sizes achieves worse performance than the iBGe methods proposed here do at the smaller sample size. The very good performance of the iBGe score at $N=400$ becomes near perfect at the larger sample size of $N=4000$ (\fref{fig:generalsim_zoom}).

\section{Biological perturbation data}

\begin{figure}[t]
  \centering
  \includegraphics[width=\textwidth]{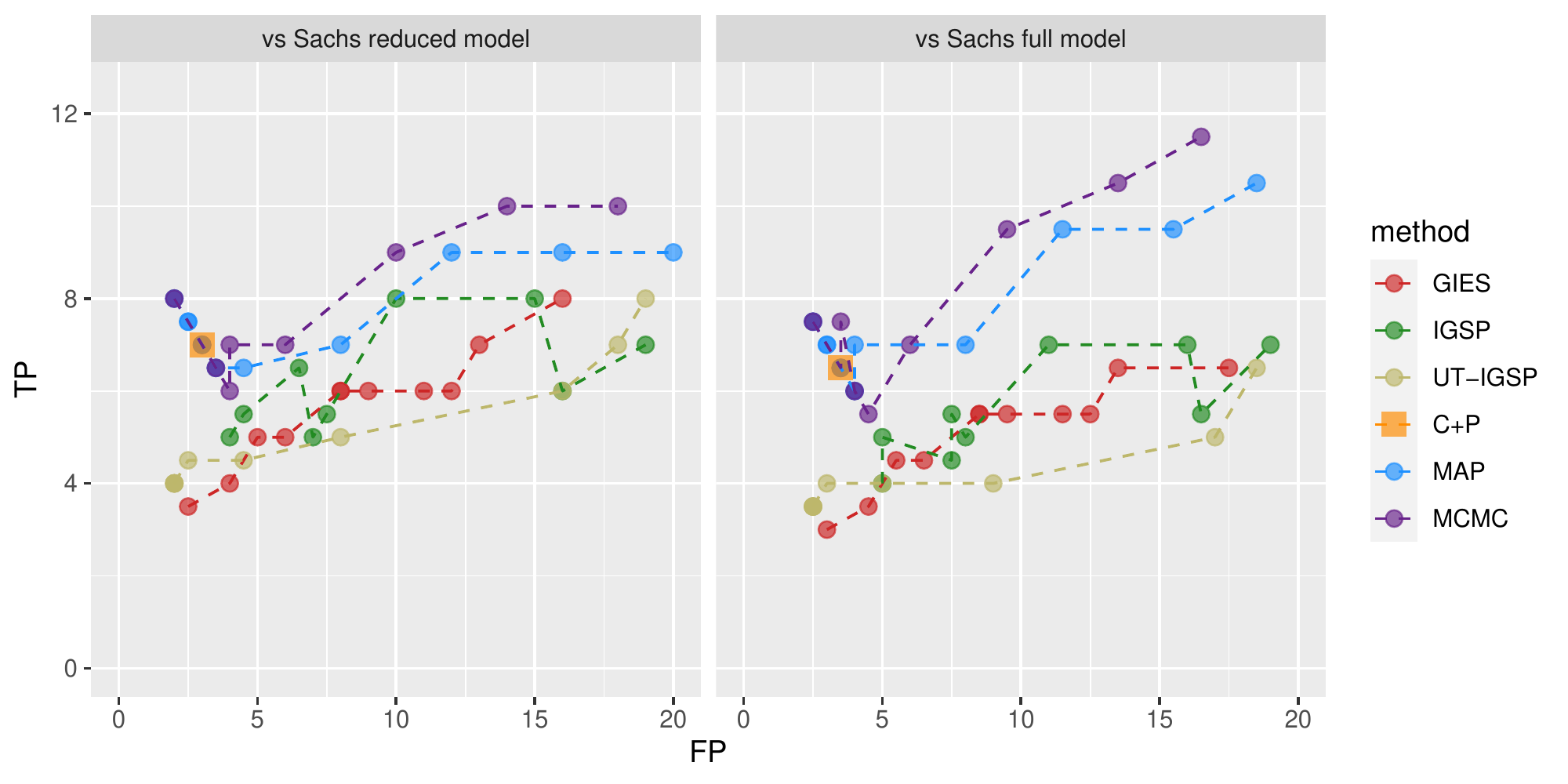}
  \caption{Performance on reconstructing the network of \citet{sachsetal05}. We compare the iBGe score input into the MAP and MCMC schemes of \pkg{BiDAG} with UT-IGSP, IGSP, GIES and the consensus network of \citet{castelletti2022network}, labelled C+P, in terms of the number of TPs and FPs defined as in \sref{sec:perf}. We compare to the canonical network of \citet{sachsetal05} both without their missing edges (reduced model) and with (full model).}
    \label{fig:sachs}
\end{figure}

To compare the iBGe approach to alternatives on real data we consider the commonly used dataset of \citet{sachsetal05}. For each T-cell in multivariate flow cytometry experiments, the amount of 11 phosphorylated molecules was measured via fluorescent readouts. The aim of the experiment was to quantify the causal relationship between these 11 nodes in the signalling pathway.  The experiments were repeated under 9 experimental conditions, of which we use the first 7 \citep[following][]{wang2017permutation, squires2020permutation}. These all contain the T-cell activator Anti-CD3/CD28 which is considered the underlying observational condition. Experiments 2--7 contain an additional agent which for experiments 3--7 directly targets a measured signalling node. The raw data is log-transformed but since the batch and experimental condition are the same and no further information on the experimental design are included in the dataset, we do not perform batch-correction as would be standard for such data.

We compare applying the iBGe score developed here to the UT-IGSP \citep{squires2020permutation} approach in terms of network recovery compared to the canonical network depicted in \citet{sachsetal05}, both with and without their missing (dashed) edges.  Since there are so many observations (5,846 in total) the prior parameters of the iBGe score make little difference. Therefore to obtain a ROC-like curve we vary a penalisation on edges to induce networks of different density instead. The results (\fref{fig:sachs}) demonstrate a clear advantage of the iBGe approach over the constraint-based UT-IGSP.

\begin{figure}[t]
  \centering
  \begin{tabular}{cc}
  \begin{tabular}[b]{c}\includegraphics[height=0.28\textwidth]{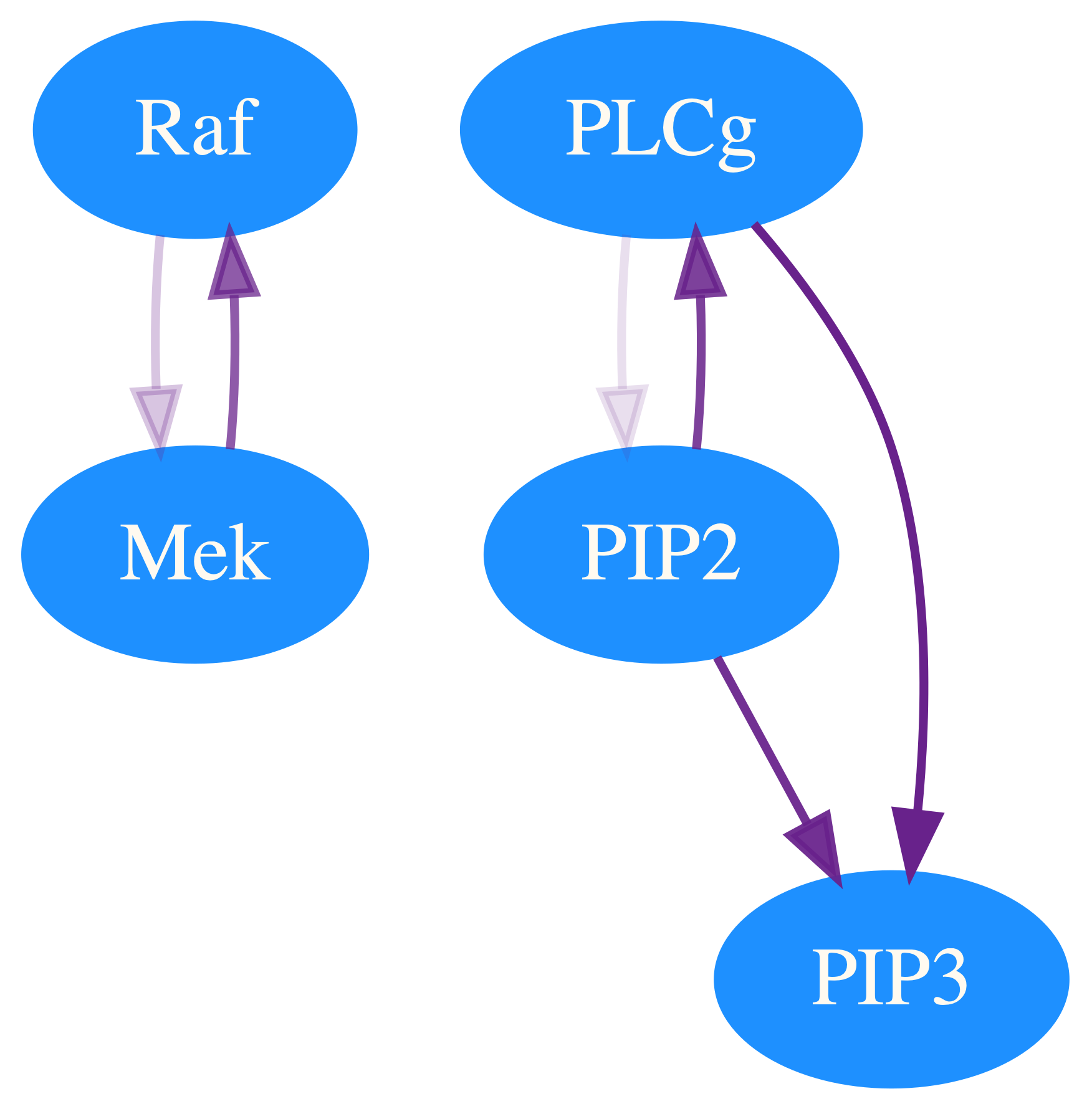} \\ 
  $\left.\right.$\\
  \includegraphics[height=0.28\textwidth]{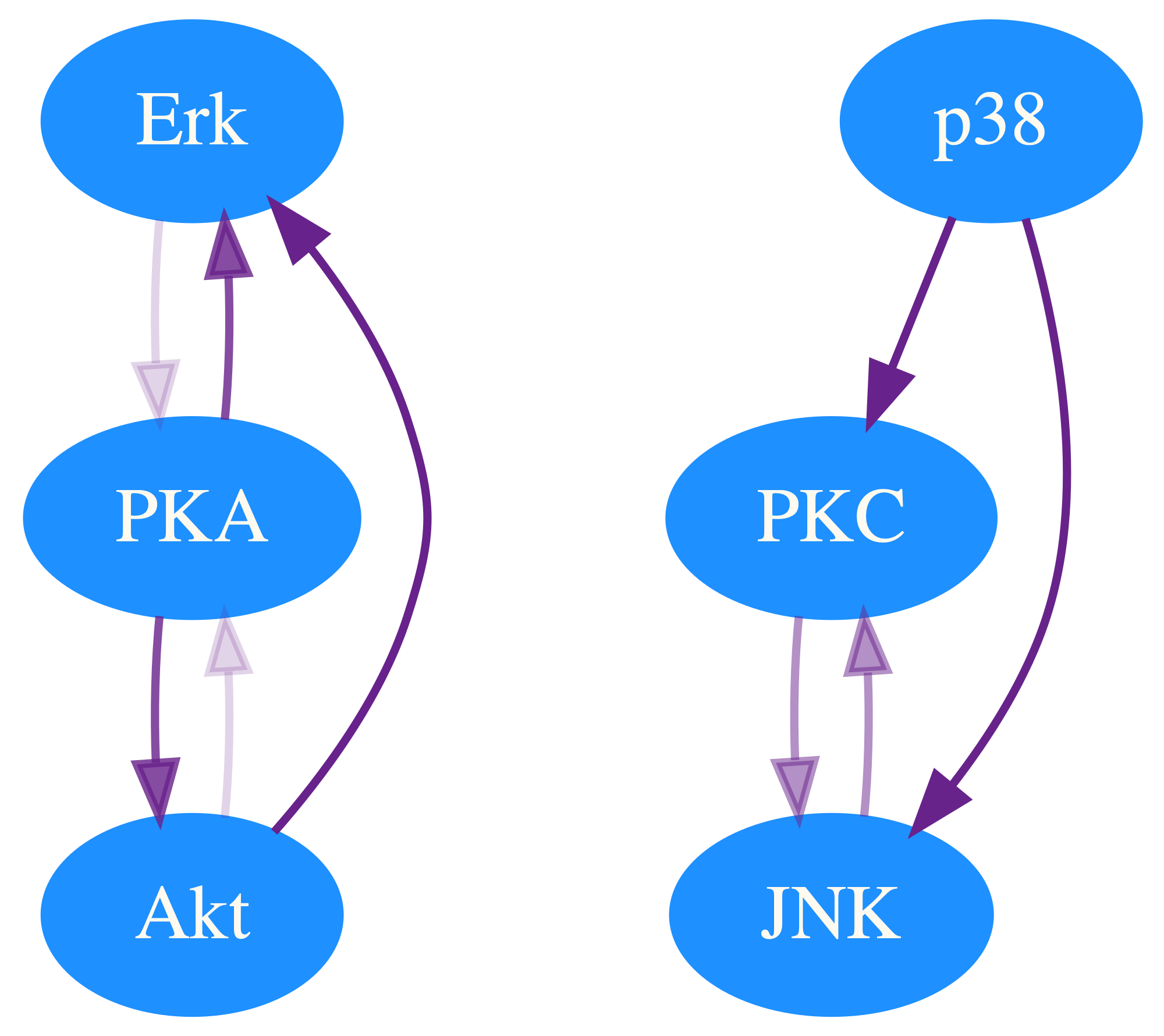}\\
  $\left.\right.$\\
  \end{tabular} & \includegraphics[width=0.63\textwidth]{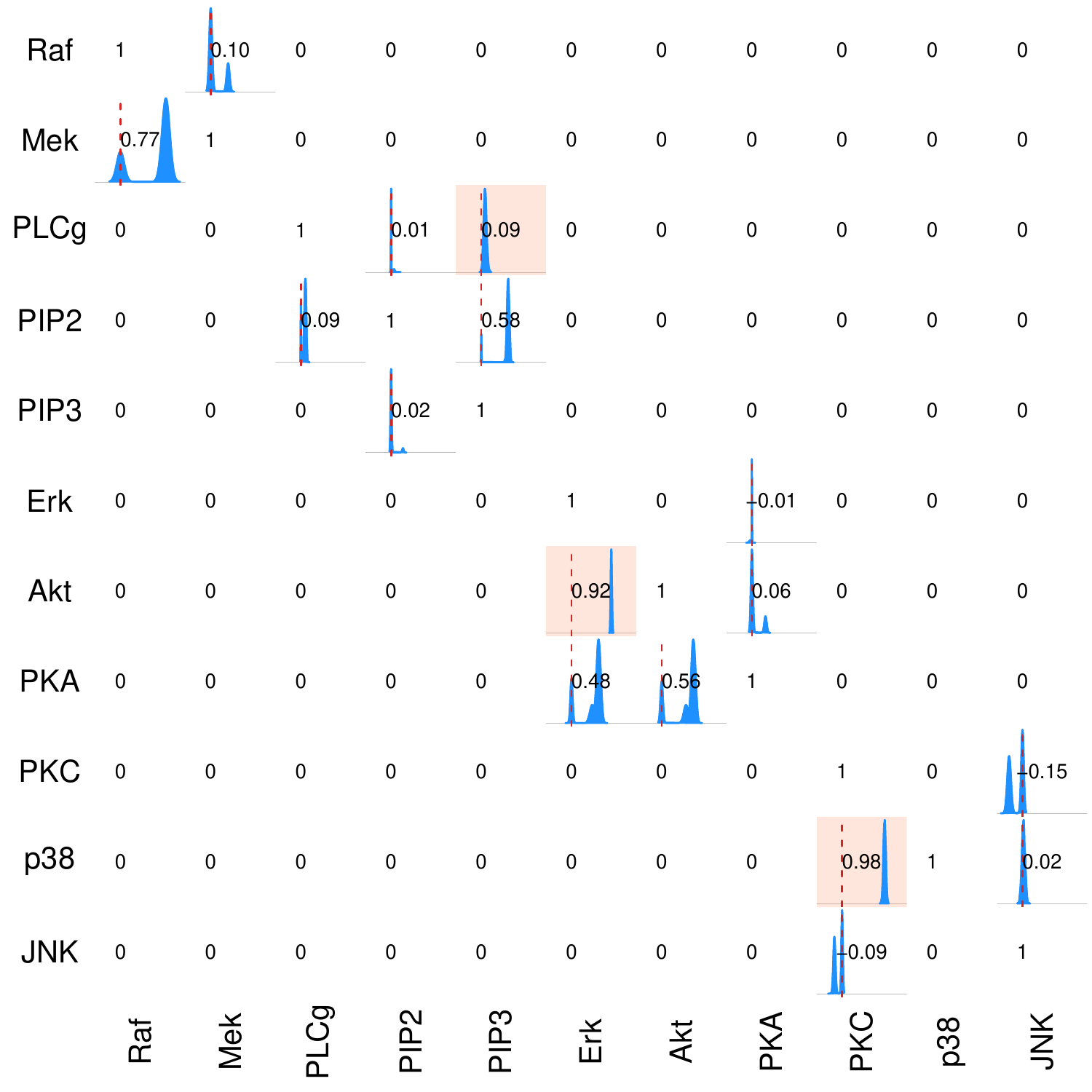} \\
  (a) & (b)
  \end{tabular}
  \caption{Summary of posterior distributions with the iBGe score on the Sachs data: (a) the posterior DAG distribution with the edge opacity corresponding to the posterior probability of the edge presence. (b) the posterior causal effect distribution with no effect indicated by the vertical dashed red lines and effects whose 95\% credible interval excludes 0 highlighted in peach.}
    \label{fig:sachseffects}
\end{figure}

When provided with the targets, IGSP \citep{wang2017permutation} and GIES \citep{hauser2012characterization} perform very similarly to each other, marginally better than UT-IGSP apart from near the origin, but still distinctly worse than the iBGe score in both the MAP and MCMC variants. The consensus network of the Bayesian approach of \citet{castelletti2022network} which handles hard interventions with unknown targets sits amongst the iBGe results (those without any edge penalisation). The iBGe however covers the more general and complex case of unknown targets with soft interventions.

We note that the canonical network considered in \citet{wang2017permutation, squires2020permutation} has an edge mistakenly reversed. Using their ground truth network simply makes all results correspondingly worse, but this does not really affect the comparative performance of \fref{fig:sachs}. Here we focused on the relative performance of different generalised approaches for continuous data, but the absolute performance of all is moderately low with a minimum SHD of 11 for the iBGe approaches, 13 for \citet{castelletti2022network}, 15 for UT-IGSP and 16 for IGSP and GIES for the reduced network with 17 edges in total. The original network of \citet{sachsetal05} was created by discretising the data, and in general network recovery for this data is quite variable, possibly due to unreliability in the ground truth network, unmeasured experimental confounding, and non-linearities and skew in the data \citep{ramsey2018fask}.    

Along with allowing us to learn the graphical structure, and its uncertainty (\fref{fig:sachseffects}a), the iBGe score allows us to further estimate the causal effects from each network (\sref{effectest}). Through Bayesian model averaging over the sampled graphs and parameters, we can then derive a sample approximation of the posterior distribution of causal effects (\fref{fig:sachseffects}b). 

\section{Conclusions}

Starting from the BGe score \citep{gh02} for graphical models with purely observational data we developed a Bayesian scoring metric for a mix of observational and interventional data. In particular, we define the score allowing interventions to be soft and unknown, so that interventions are not necessarily structural and may affect the strength of a relationship, while the targets may be unknown.  For discrete data, we may view soft interventions \citep{eaton2007exact} as interactions, and we developed a model for continous data by taking the analogy over to the continuous case. By further leveraging the connections between the SEM and matrix parametrisation of the BGe score, we could define the interventional BGe (iBGe) score as a natural combination of BGe scores over experimental conditions. The novel framework covers the case of soft interventions, while handling uncertainty in the targeting by also learning the connections between the interventions and the observational nodes.

The iBGe score we derived is easy to compute and include in score-based algorithms, allowing their easy adaptation to mixed data with unknown and soft interventions. The highlight of the BGe score, however, is that it is Bayesian so that it can further be used for MCMC and other Bayesian approaches for model averaging over DAGs. This property carries over to the iBGe score, so that it can be used with sampling methods for structure learning, such as order \citep{fk03} or partition MCMC \citep{km17}, as we do by interfacing the score with a hybrid approach \citep{ksm22} implemented in the \pkg{BiDAG} package \citep{skmb21}. The iBGe approach, especially combined with such state-of-the-art hybrid inference \citep{ksm22}, outperforms current alternatives like UT-IGSP \citep{squires2020permutation} in simulation studies and for real data. 

The Bayesian approach to causal structure learning accomplishes some important analysis tasks: quantifying the uncertainty in the network structure, characterising the uncertainty in the parameter distributions and automatically propagating both into the downstream analyses of intervention effects \citep{moffa2017using, kuipers2019links, moffa2021longitudinal}. As with the BGe score for purely observational data \citep{viinikka2020towards}, the iBGe score now enables the same Bayesian inference of causal effects for mixed observational and interventional data.

The biological perturbation data of \citet{sachsetal05}, displays non-Gaussian skewness and possible non-linearity breaking the underlying linear-Gaussian assumptions of the BGe and iBGe scores. Although constraint-based methods can relatively easily change their conditional independence tests, building a marginalisable likelihood like the BGe suitable for Bayesian analyses in the presence of non-linearity and non-Gaussianity seems more challenging. For scores developed for non-Gaussian observational data, however, we can expect that the approach developed here will allow for a direct extension to handle data with soft unknown interventions.

\vskip 0.2in
\bibliography{iBGe}

\clearpage

\appendix

\section*{Supplementary Material}
\setcounter{section}{0} 

\renewcommand\thefigure{S\arabic{figure}}
\setcounter{figure}{0} 

\section{Simulation study with perfect interventions}
\label{app:perfect_sims}

\begin{figure}[t]
  \centering
  \includegraphics[width=\textwidth]{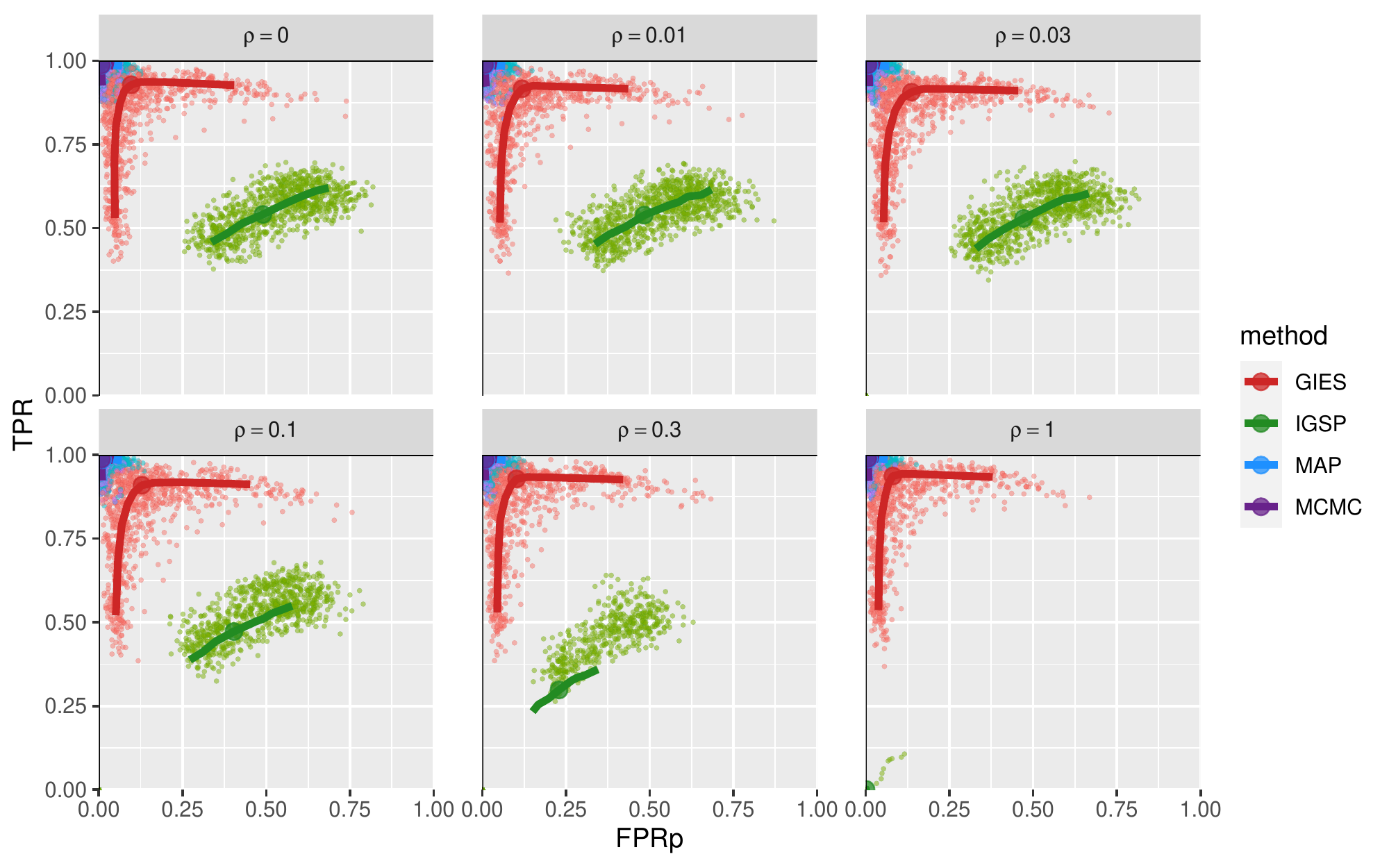}
  \caption{With perfect and known interventions, a comparison of the iBGe score input into the MAP and MCMC consensus scheme of \pkg{BiDAG} to IGSP and GIES, as the fraction of interventional data $\rho$ increases. Each point is a single repetition for a single parameter value, while the thicker lines show the average behaviour for each parameter value with the larger dot placed at  $10^{5}\alpha = 1 = 10\am$, and $\lambda = 2.3$. IGSP often returns the empty DAG at (0,0) with more limited amounts of purely observational data at larger $\rho$, leading to the divergence between its cloud of dots and average line.}
  \label{fig:perfectsim}
\end{figure}

With perfect interventions, we can additionally compare to GIES \citep[Greedy Interventional Equivalence Search;][]{hauser2012characterization} and IGSP \citep[Interventional Greedy Sparsest Permutation;][]{wang2017permutation}, the precursor of UT-IGSP \citep{squires2020permutation} with known targets. We follow the same simulation strategy as in the main text, select 10 nodes randomly to be targets, and fix a fraction of the data $\rho = (0, 0.01, 0.03, 0.1, 0.3, 1)$ to be interventional. This covers the range from fully observational to fully interventional. Amongst the interventional data, we randomly select the target for each observation. By default we did not standardise the data, since then IGSP failed to perform. In the comparison, we use the following range of penalisation parameters for GIES
\begin{equation}
\lambda = (0.607, 0.847, 1.18, 1.65, 2.3, 3.21, 4.48, 6.26, 8.74)
\end{equation}

\begin{figure}[t]
  \centering
  \includegraphics[width=\textwidth]{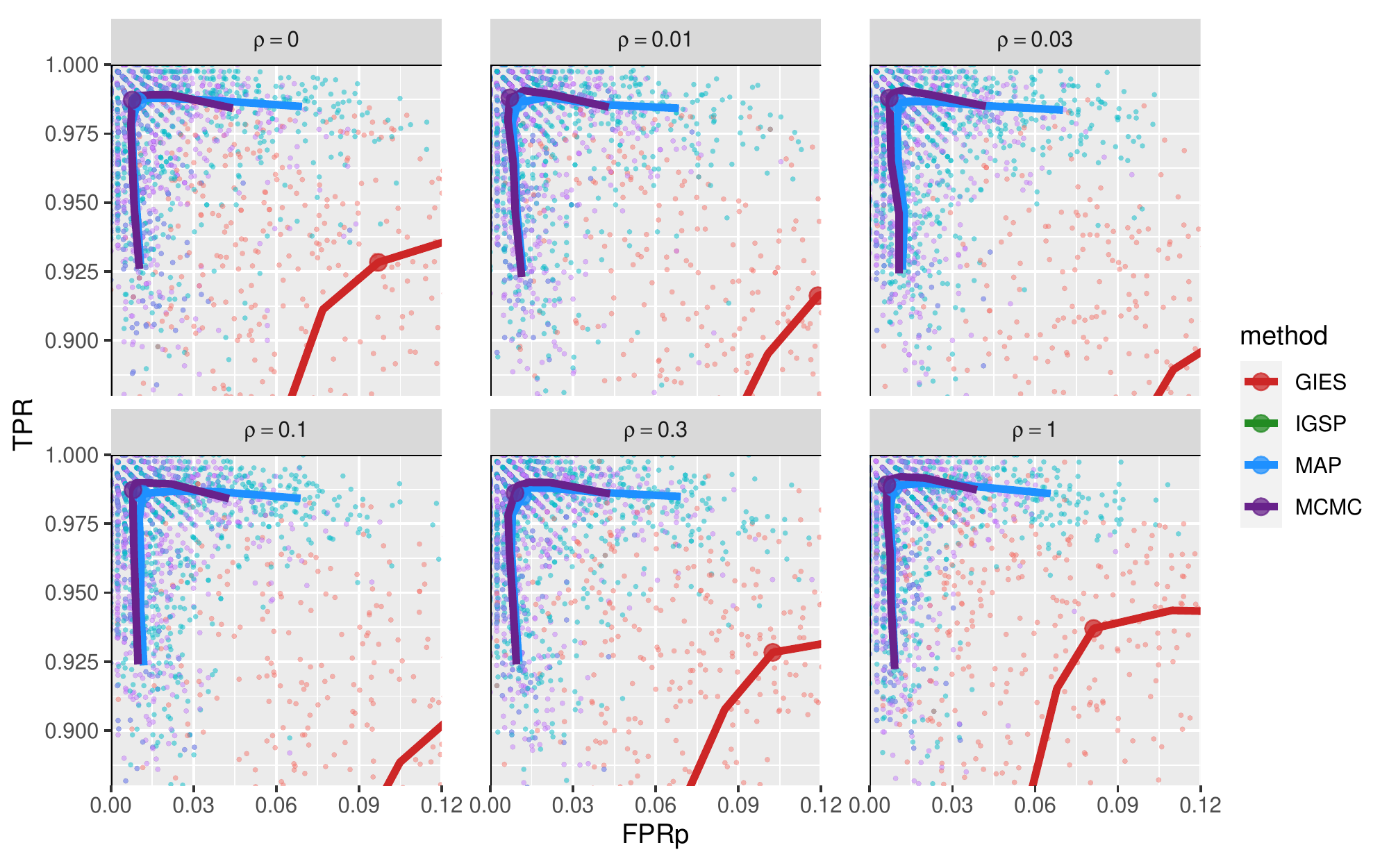}
  \caption{Zoom into the top left of \fref{fig:perfectsim} to better compare the iBGe score input into the MAP and MCMC inference schemes of \pkg{BiDAG} to GIES.}
  \label{fig:perfectsim_zoom}
\end{figure}

The constraint-based algorithm of IGSP performs relatively poorly in this setting (\fref{fig:perfectsim}), especially as it seems to require more observational data, often returning the empty DAG when there is too much interventional data. The iBGe score and GIES are more robust, with a clear strong advantage to using the iBGe score over GIES in terms of performance (\fref{fig:perfectsim_zoom}).

\begin{figure}[t]
  \centering
  \includegraphics[width=0.75\textwidth]{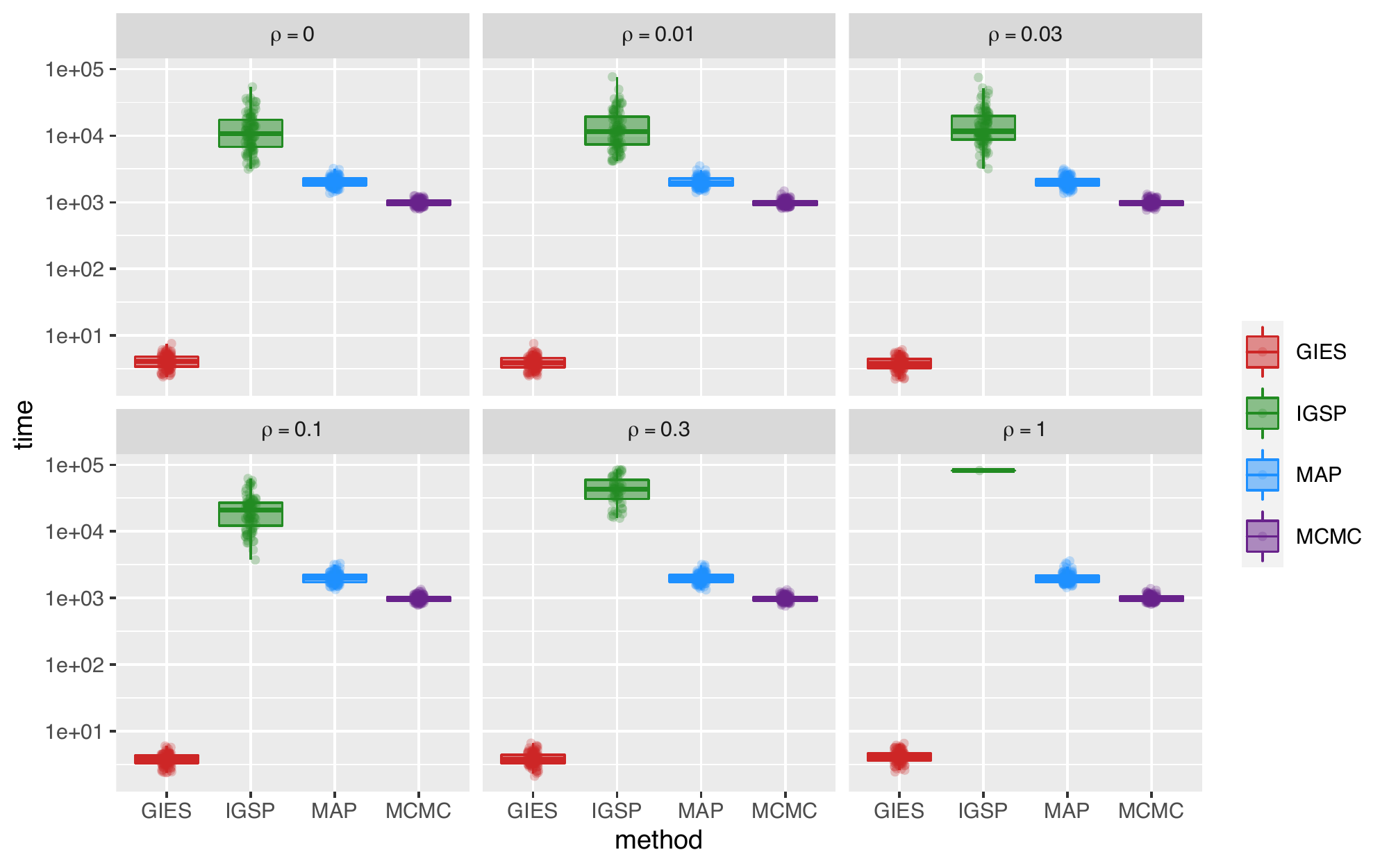}
  \caption{The time taken to learn a network with GIES, IGSP, and the MAP and MCMC consensus graphs of \pkg{BiDAG} running with the iBGe score for known perfect interventions. The MCMC scheme requires the MAP steps to be run first and its times are the additional time for the sampling.}
  \label{fig:perfectsim_time}
\end{figure}

Timewise (\fref{fig:perfectsim_time}), GIES is much faster than the sampling-based inference schemes of \pkg{BiDAG}, in line with results for purely observational data \citep{ksm22}, but at the cost of worse performance. IGSP is slower still, but this may depend heavily on the implementation as the python-based UT-IGSP runs notably faster.

\begin{figure}[t]
  \centering
  \includegraphics[width=0.75\textwidth]{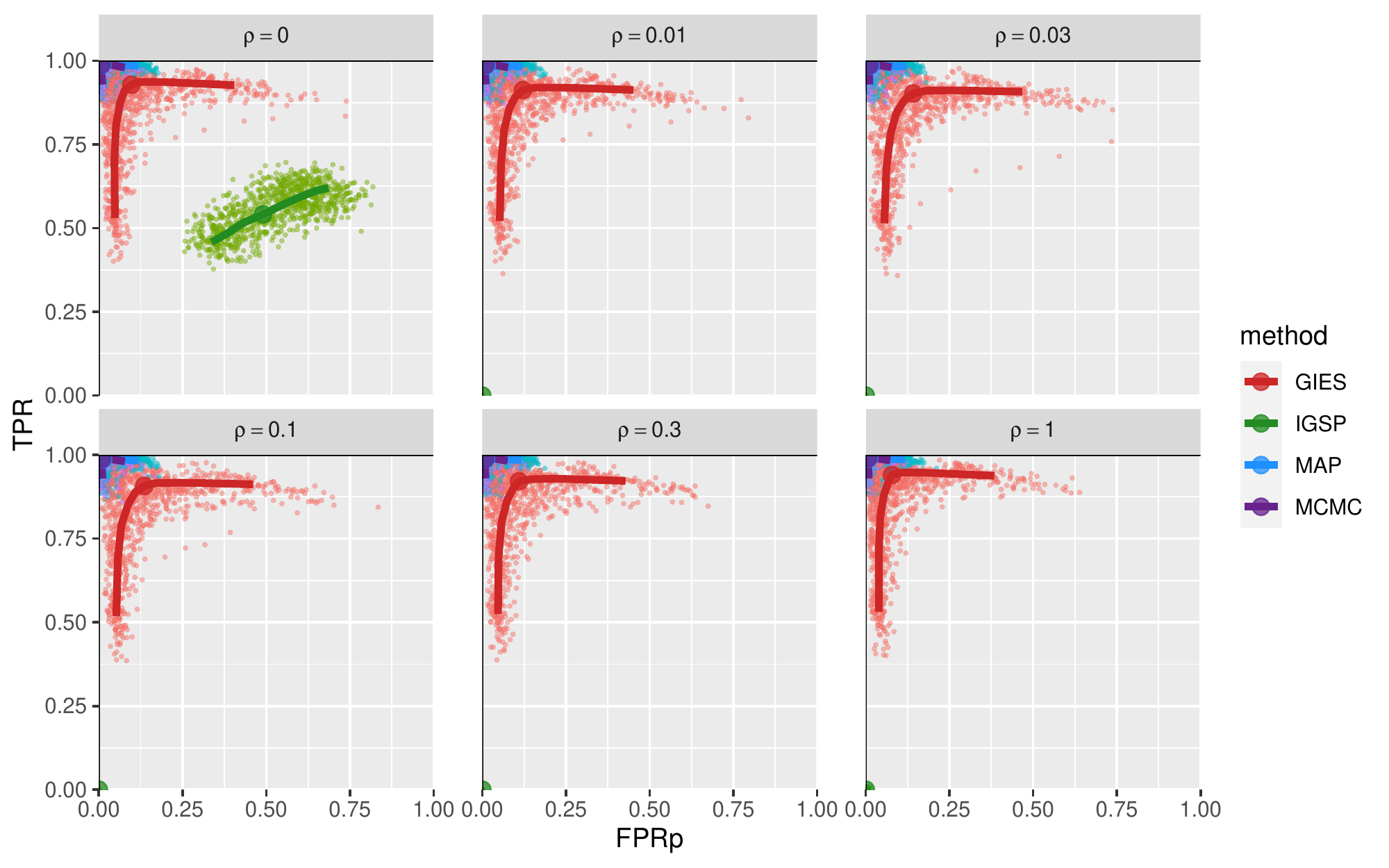}
  \caption{Comparison for known perfect interventions, as in \fref{fig:perfectsim}, but where the data has been standardised. This highlights that IGSP fails in this setting while the iBGe score (MAP and MCMC) and GIES are robust.}
  \label{fig:perfectsim_scaled}
\end{figure}

For completeness we include the results when we standardised the data (\fref{fig:perfectsim_scaled}) to remove the possibility of artificially using the scale of the data to improve performance \citep{reisach2021beware}. As expected, the performance of the iBGe score and GIES are relatively unchanged. However, IGSP only returns the empty DAG for $\rho>0$ and fails to learn meaningful DAGs.

\end{document}